\documentclass[12pt]{article}
\pdfoutput=1

\usepackage{amsmath,graphicx}
\usepackage{epsf}
\usepackage{graphicx,epsfig}
\usepackage{amsfonts}
\usepackage{amssymb}
\usepackage[nosort]{cite}
\usepackage{setspace}

\def\CC{{\cal C}}

\def\CH{{\cal H}}
\def\CJ{{\cal J}}

\def\CO{{\cal O}}

\def\half{\frac{1}{2}}

\makeatletter
\renewcommand\section{\@startsection {section}{1}{\z@}%
                                 {-3.5ex \@plus -1ex \@minus -.2ex}%
                                   {2.3ex \@plus.2ex}%
                                   {\normalfont\large\bfseries}}
\renewcommand\subsection{\@startsection{subsection}{2}{\z@}%
                                   {-3.25ex\@plus -1ex \@minus -.2ex}%
                                     {1.5ex \@plus .2ex}%
                                     {\normalfont\bfseries}}
\renewcommand\subsubsection{\@startsection{subsubsection}{3}{\z@}%
                                   {-3.25ex\@plus -1ex \@minus -.2ex}%
                                     {1.5ex \@plus .2ex}%
                                     {\normalfont\itshape}}
\makeatother

\newcommand{\Letter}{
\setlength{\textwidth}{16.5cm}
   \setlength{\textheight}{22.6cm}
    \hoffset=-0.5in
\voffset=-2.1cm }

\Letter

\begin{document}
\newcommand{\be}{\begin{equation}}
\newcommand{\ee}{\end{equation}}
\newcommand{\bea}{\begin{eqnarray}}
\newcommand{\eea}{\end{eqnarray}}
\newcommand{\barr}{\begin{array}}
\newcommand{\earr}{\end{array}}

\thispagestyle{empty}
\begin{flushright}
\end{flushright}

\vspace*{0.3in}
\begin{spacing}{1.1}

\begin{center}
{\large \bf Quasi-Single Field Inflation with Large Mass}

\vspace*{0.5in} {Xingang Chen${}^1$ and Yi Wang${}^2$}
\\[.3in]
\textit{
${}^1$ Center for Theoretical Cosmology,\\
Department of Applied Mathematics and Theoretical Physics,\\
University of Cambridge, Cambridge CB3 0WA, UK \\
\smallskip
${}^2$ Physics Department, McGill University, \\
Montreal, QC, H3A 2T8, Canada} \\[0.3in]
\end{center}

\begin{center}
{\bf
Abstract}
\end{center}
\noindent
We study the effect of massive isocurvaton on density perturbations in quasi-single field inflation models, when the mass of the isocurvaton $M$ becomes larger than the order of the Hubble parameter $H$. We analytically compute the correction to the power spectrum, leading order in coupling but exact for all values of mass. This verifies the previous numerical results for the range $0<M<3H/2$ and shows that, in the large mass limit, the correction is of order $H^2/M^2$.

\vfill

\newpage
\setcounter{page}{1}


\newpage

\section{Introduction}

Massive fields are present during inflation, as a consequence of realistic model-building and UV completion. In terms of the low energy effective field theory, the inflaton travels along a trajectory determined by the minima of some complicated potential landscape spanned by a large number of fields. The scalar fields orthogonal to this trajectory, the isocurvatons, have a rich spectrum. In inflation scenarios from string theory and supergravity, the lightest of them start from order $H$, the Hubble parameter during inflation. This value is determined by the coupling of the scalar fields to the spacetime curvature and stabilized by supersymmetry. Lighter values require fine-tuning or extra symmetry. Those inflation models with isocurvaton mass of order $\CO(H)$ are called the quasi-single field inflation models \cite{Chen:2009we,Chen:2009zp,Baumann:2011nk,Emiliano:2012,Norena:2012}. The simplest version of such models has been used to show the distinctive effect of these massive isocurvatons on the primordial non-Gaussianities. In this paper, we would like to consider the model for a different purpose. We will investigate how exactly the massive field decouples when its mass becomes larger than $\CO(H)$.

In addition to the isocurvatons with mass of order $\CO(H)$, there are certainly more fields that have masses larger than $\CO(H)$. Intuitively the quantum fluctuations of such fields should be less important to the low energy effective theory description. The purpose of this paper is to make this statement more precise and quantitative in the context of the quasi-single field inflation, which represents a quite general class of models in model-building. The effects of heavy fields have been studied in different models previously \cite{Kaloper:2002uj,Jackson:2010cw}. It has been argued that their corrections to the power spectrum are proportional to powers of $H/M$ for large $M/H$, and the specific power is model-dependent. In the simplest model of quasi-single field inflation studied in \cite{Chen:2009we,Chen:2009zp}, the massive field couples when the inflaton trajectory turns, and the coupling appears in the kinetic term.
We would like to compute the effect of this massive field on power spectrum for all values of $M$. Generally speaking, obtaining such a correction in an analytically closed form from the first principles is difficult, and such an example was not available so far. This is because the in-in formalism involves multiple integrations of the product of the mode functions, and the mode function for the massive field contains a special function. It turns out that the model example studied in \cite{Chen:2009we,Chen:2009zp} is just simple enough for us to achieve this goal here, thus providing a solid, and reasonably general case on how precisely we mean that massive fields can be integrated out.

Before the detailed computation, we may have two different guesses what the final result should be qualitatively. The first possibility is by analogy to the thermal field theory, in which the contributions of massive states to correlation functions are exponentially suppressed by a Boltzmann factor if the mass is much higher than the temperature. Here in de Sitter space, we have a Gibbons-Hawking temperature $T_{GH}=H/2\pi$, so the corresponding Boltzmann factor is $e^{-2\pi M/H}$. The second is by analogy to the scattering process in particle physics, in which the contribution of massive states to the scattering amplitude is power-law suppressed if the mass is much higher than the energy scale of the process.
Interestingly as we will see, both types of contributions appear in different terms in the in-in formalism, and the latter is more general. Consequently, for large $M/H$, the power-law suppression dominates.

\section{The model and a summary of the results}
\label{sec:model-summary-result}
\setcounter{equation}{0}

For convenience we review the model and summarize the main results in this section.

We consider the constant turn case of a two-field quasi-single field inflation model \cite{Chen:2009we,Chen:2009zp},
\bea
S = \int d^4x \sqrt{-g} \left[
-\half (\tilde R + \sigma)^2 g^{\mu\nu} \partial_\mu \theta \partial_\nu \theta
- \half g^{\mu\nu} \partial_\mu \sigma \partial_\nu \sigma
- V_{\rm sr}(\theta) - V(\sigma) \right] ~.
\eea
This describes an inflaton that is moving around a circle in field space with radius $\tilde R$, and at the same time rolling down the potential. The $\theta$ is the angular coordinate, and $\sigma$ is the radial coordinate. In the angular direction, the potential is a slow-roll potential $V_{\rm sr}(\theta)$; while in the radial direction the potential is lifted around the effective minimum $\sigma_0$ so the isocurvaton is massive.
The massive field not only determines the bending of the trajectory classically, but also influences the fluctuations of the inflaton field quantum-mechanically.
In the infinite mass limit $M \to \infty$, this Lagrangian reduces to a single field model, with radius fixed at $R=\tilde R + \sigma_0$ and the inflaton field being $R\theta$. The purpose of this paper is to demonstrate how this limit is reached and what the dependence of power spectrum on the mass $M$ is for a finite $M$.

After perturbing the two fields, we obtain the following kinematic Hamiltonian density
\bea
\CH_0 = a^3 \left[ \half R^2 \dot {\delta\theta_I}^2 +
  \frac{R^2}{2a^2}
  (\partial_i \delta\theta_I)^2
+ \half \dot{\delta\sigma_I}^2 + \frac{1}{2a^2} (\partial_i
\delta\sigma_I)^2 + \half M^2 \delta \sigma_I^2
\right] ~,
\label{H0}
\eea
and the interaction Hamiltonian density
\bea
\CH^I_2 &=& - 2R \dot\theta_0~ a^3 \delta\sigma_I \dot{\delta\theta_I} ~.
\label{CH2}
\eea
We have defined $M$ to be the effective mass appearing in the Hamiltonian, $M^2 = V''(\sigma_0) + 3 \dot\theta_0^2$, and $\dot\theta_0$ is the constant turning angular velocity.
We are only interested in the quadratic part of the interaction Hamiltonian in this paper. The label ``$I$" denotes that the fields are in the interaction picture.

\begin{figure}[htbp]
\centering
\includegraphics[width=0.3\textwidth]{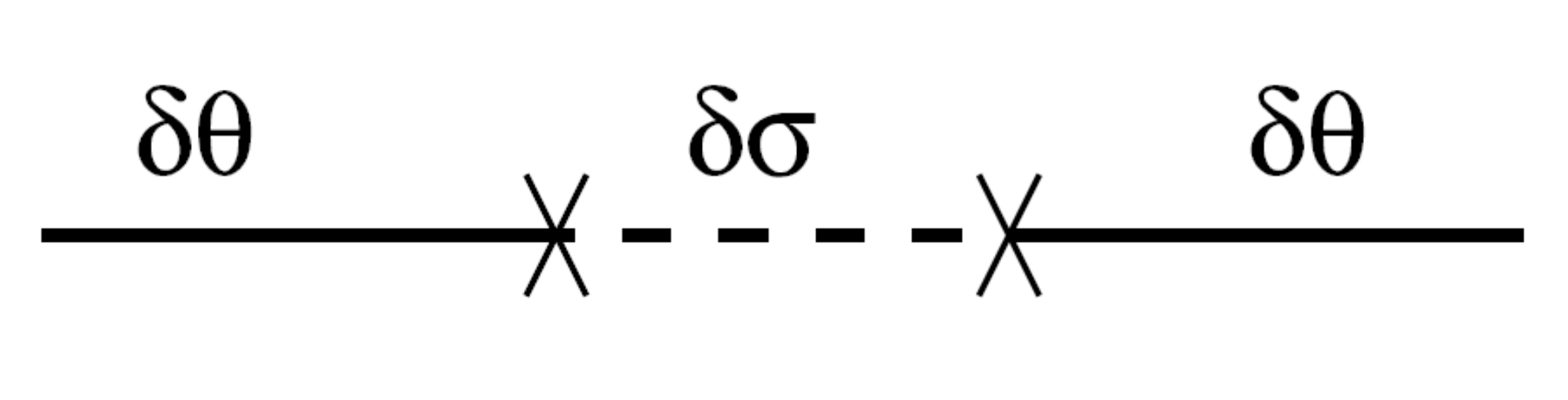}
\caption{\label{Fig:Fdiagram-2pt} The correction to the leading power spectrum.}
\end{figure}

The quadratic interaction Hamiltonian introduces a coupling between the massive isocurvaton and the inflaton, and the strength of the coupling is constant and proportional to $\dot\theta_0$. Due to the constant turn assumption, we are excluding various phenomena associated with sharp features in turning trajectory \cite{Achucarro:2010da,Chen:2011zf,Chen:2011tu,Shiu:2011qw,Achucarro:2012sm,Avgoustidis:2012yc}.
The leading order correction to the power spectrum comes from the interaction illustrated by the diagram Fig.~\ref{Fig:Fdiagram-2pt}, and the corresponding terms in the in-in formalism (see \cite{Weinberg:2005vy,Chen:2010xka} for review) are
\bea
\langle \delta\theta^2 \rangle
&\supset&
\int_{t_0}^t d\tilde t_1 \int_{t_0}^t dt_1
\langle 0| H_I(\tilde t_1) ~\delta\theta_I^2~ H_I(t_1) |0\rangle
\nonumber \\
&-& 2 ~{\rm Re} \left[
\int_{t_0}^t dt_1 \int_{t_0}^{t_1} dt_2
\langle 0| \delta\theta_I^2~ H_I(t_1) H_I(t_2) |0\rangle \right] ~.
\label{2pt_correction}
\eea
We assume the Bunch-Davies vacuum for both mode functions, $\delta\theta_I$ and $\delta\sigma_I$.
The final power spectrum is given by \cite{Chen:2009zp}
\begin{align}
P_\zeta=&
\frac{H^4}{4\pi^2 R^2 \dot\theta_0^2}\left[
1+8 {\cal C} \left( \frac{\dot\theta_0}{H} \right)^2 \right] ~.
\label{power}
\end{align}
In \cite{Chen:2009zp}, the coefficient $\CC$ is computed numerically for $M/H<3/2$. In this work we will compute this coefficient for all $M$ analytically.
The coefficient $\CC$ contains two contributions
\begin{align}
\CC &= \CC_1 + \CC_2 ~,
\\
\CC_1 &\equiv \frac{\pi}{8}e^{-\mu\pi}
  \left|
    \int_0^\infty dx_1 ~
    x_1^{-1/2}H_{i\mu}^{(1)}(x_1)e^{ix_1}
  \right|^2 ~,
\label{CC1} \\
\mathcal{C}_2 &\equiv -\frac{\pi}{4} ~ e^{-\mu\pi}~\mathrm{Re}
  \left\{
    \int_0^\infty dx_1 ~
      x_1^{-1/2}H_{i\mu}^{(1)}(x_1)e^{-ix_1} \int_{x_1}^\infty dx_2 ~
      x_2^{-1/2}\left[H_{i\mu}^{(1)}(x_2)\right]^* e^{-ix_2}
  \right\} ~,
\label{CC2}
\end{align}
coming from the two terms in (\ref{2pt_correction}) respectively.
The variable $x_i \equiv -i k_1 \tau_i$ where $\tau_i$ is the conformal time.
$H^{(1)}_{i\mu}$ is the Hankel function of the first kind. The parameter $\mu$ is defined to be
\bea
\mu \equiv \sqrt{ \frac{M^2}{H^2} - \frac{9}{4} } ~.
\eea
It can be either real or imaginary. For real $\mu$, the isocurvaton is heavier, $M\ge 3H/2$, which is the main interest of this paper. For imaginary $\mu$, this is related to the parameter $\nu$ in \cite{Chen:2009we,Chen:2009zp} by $\mu=-i\nu$, and the isocurvaton has a mass $0 \le M <3H/2$. Also note that, for the latter case, the factors $e^{-\mu\pi}$ in (\ref{CC1}) and (\ref{CC2}) disappear.

The first term is a product of two integrals without time-ordering, one from the interacting vacuum and another from its complex conjugate. It turns out that in the large $\mu$ limit, this term contains a Boltzmann factor $\CC_1 \to \pi^2 e^{-2\pi M/H}$. The second term involves a time-ordering double integral, similar to the form of scattering amplitude in the in-out formalism. In the large $\mu$ limit, the suppression turns out to be power-law $\CC_2 \to H^2/(4M^2)$. We expect the power-law to be the more generic behavior in the in-in formalism. Therefore overall $\CC \to H^2/(4M^2)$.
This verifies the result from the semi-classical approximation in the large $M$ limit \cite{Tolley:2009fg,Achucarro:2010da}.

\begin{figure}[htbp]
\centering
\includegraphics[width=0.6\textwidth]{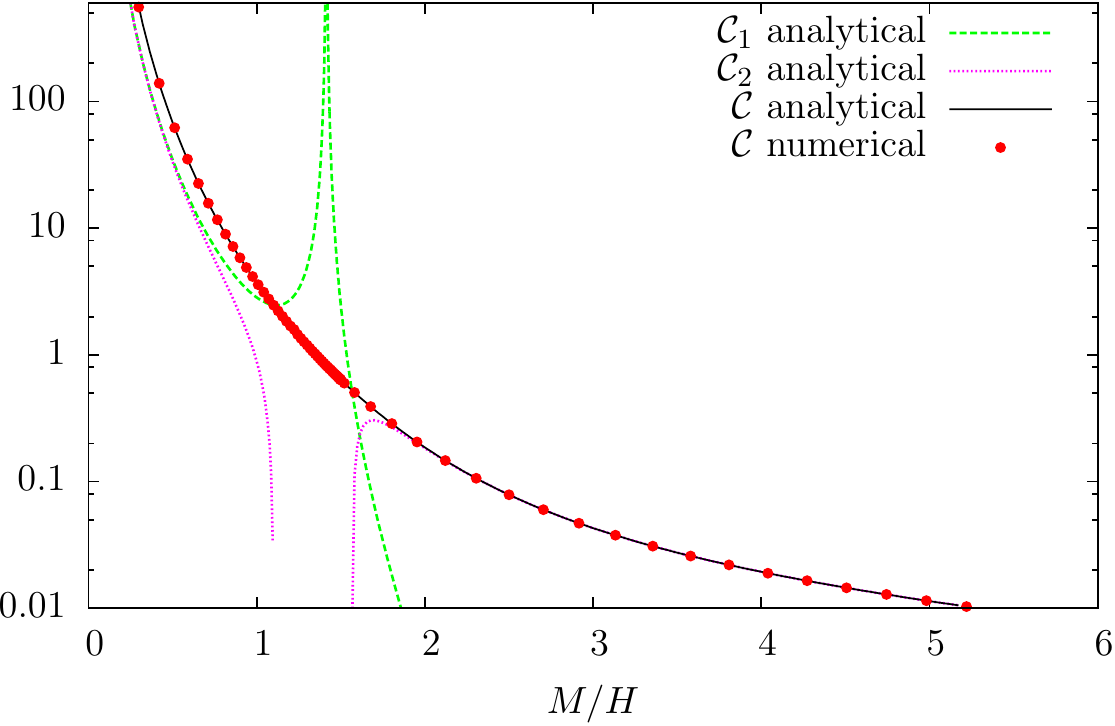}
\caption{\label{fig:MHlog-crop} The final result $\CC$ (solid lines), $\CC_1$ and $\CC_2$ (dashed lines) as a function of $M/H$ for both $M\sim H$ and $M\gg H$. The dots are numerical results.}
\end{figure}

The complete results with arbitrary $\mu$ (i.e. arbitrary isocurvaton mass) are
\bea
\mathcal{C}_1 &=& \frac{\pi^2}{4\cosh^2(\pi\mu)} ~,
\\
\CC_2 &=& \frac{e^{\pi\mu}}{16 \sinh\pi\mu}
    \mathrm{Re}\left[
      \psi^{(1)}\left(\frac{3}{4}+\frac{i\mu}{2}\right)
      -\psi^{(1)}\left(\frac{1}{4}+\frac{i\mu}{2}\right)
    \right]
    \nonumber \\
&-& \frac{e^{-\pi\mu}}{16 \sinh\pi\mu}
    \mathrm{Re}\left[
      \psi^{(1)}\left(\frac{3}{4}-\frac{i\mu}{2}\right)
      -\psi^{(1)}\left(\frac{1}{4}-\frac{i\mu}{2}\right)
    \right] ~,
\eea
where $\psi^{(1)}$ is the polygamma function,
\begin{align}
  \psi^{(1)}(z)\equiv \frac{d^2 \ln \Gamma(z)}{dz^2} ~.
\end{align}
We plot the coefficient $\CC$ in Fig.~\ref{fig:MHlog-crop}.

If we are only interested in the behavior in the large $M/H$ limit, a simpler \textit{IR expansion method} may be used. In this method, we replace the UV part of the mode function of the massive field by its IR approximation, which is actually valid between the super-Hubble scale $k\tau \to 0$ and the sub-Hubble scale $k\tau \sim -\sqrt{M/H}$. This produces the correct answer \textit{if} the difference between the replaced part and the original one is much smaller than the final result. We will also show that the dominant contribution from these heavy fields comes from between of order $\ln \mu$ e-folds before the Hubble crossing point $k\tau =-1$, and a few e-folds after Hubble crossing.
Therefore, the main results of this paper also apply to the non-constant-turn case, as long as the variation of the angular velocity and mass is not very fast, $\ddot\theta/(H\dot\theta) \ll 1$ and $\dot M/(H M) \ll 1$.

The rest of the paper is organized as follows: In the following two sections, Sec.~\ref{sec:c1} and \ref{sec:c2}, we give the computational details and tricks for the two integrals. In Sec.~\ref{Sec:Analytical_continuation}, we analytically continue the final result to imaginary $\mu$, and show that they match the expected properties and the numerical results in \cite{Chen:2009zp}. In Sec.~\ref{Sec:IR_Approx}, we give a simpler IR expansion method that can be used to reproduce the analytical result in the large $M/H$ limit. In Sec.~\ref{Sec:Truncating}, we investigate where the dominant contribution of the massive field comes from during its evolution by examining the integrals as a function of an IR cutoff $x_c$. In the Appendix, we describe a numerical method for checking the integrals.

\section{The integral without time-ordering}
\label{sec:c1}
\setcounter{equation}{0}

We start with the simpler integral,
\begin{align}\label{eq:c1}
  \mathcal{C}_1 \equiv \frac{\pi}{8}e^{-\mu\pi}
  \left|
    \int_0^\infty dx_1 ~
    x_1^{-1/2}H_{i\mu}^{(1)}(x_1)e^{ix_1}
  \right|^2 ~.
\end{align}

\begin{figure}[htbp]
\centering
\includegraphics[height=0.34\textheight]{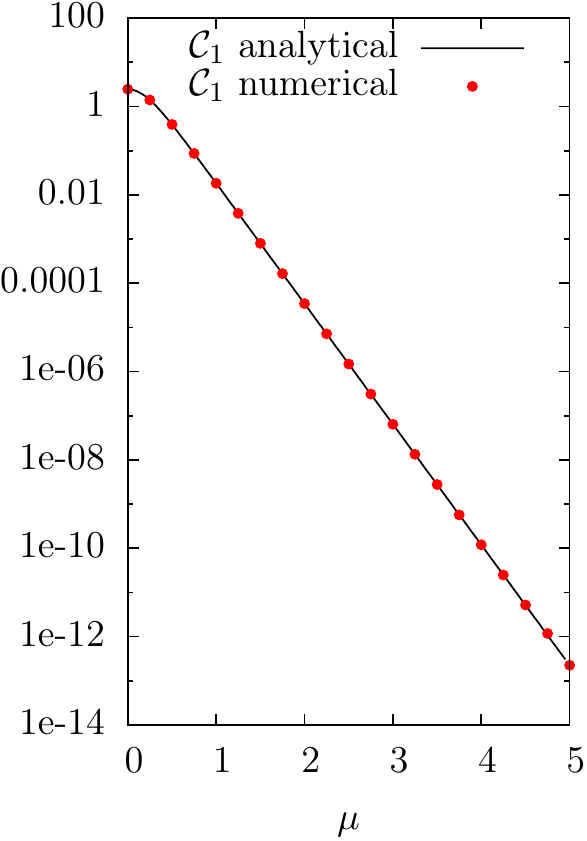}
\includegraphics[height=0.34\textheight]{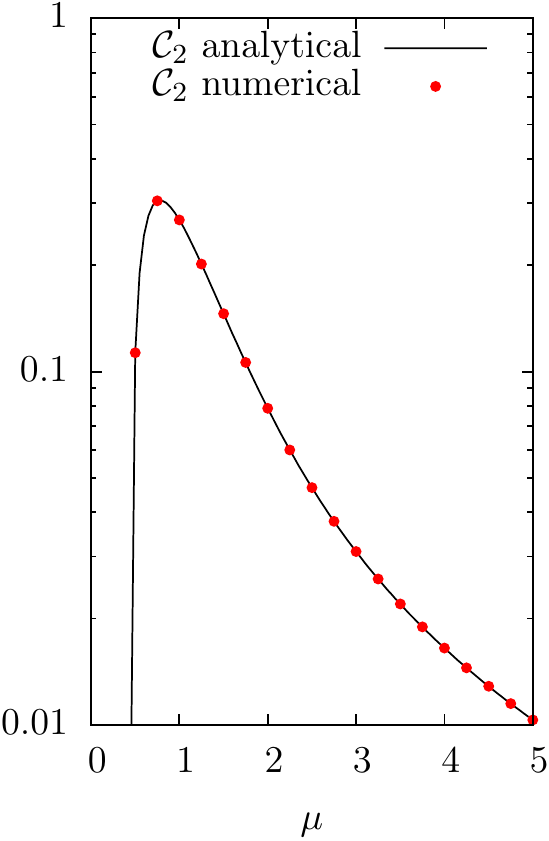}
\includegraphics[height=0.34\textheight]{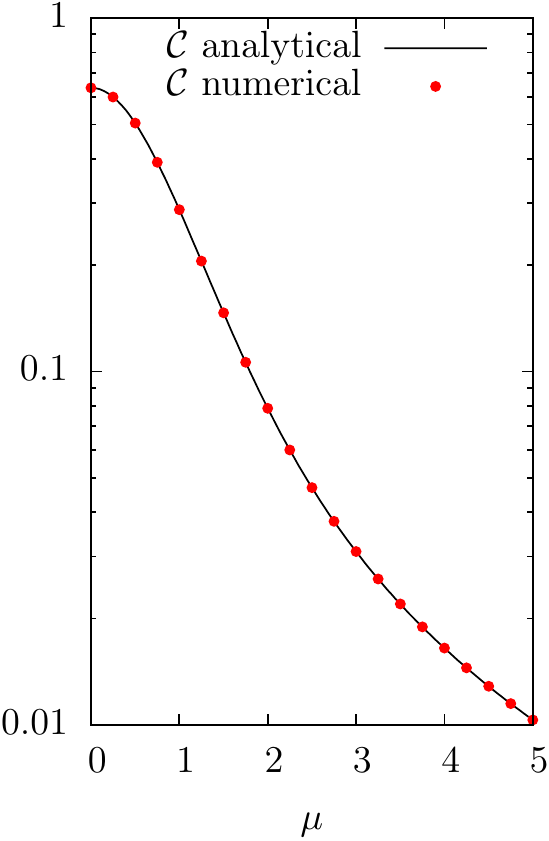}
\caption{\label{fig:num} Comparison of analytical and numerical results. The left, middle, right panels are analytical and numerical plots for $\mathcal{C}_1$, $\mathcal{C}_2$ and $\mathcal{C}\equiv \mathcal{C}_1 + \mathcal{C}_2$ respectively. It is clear from the plot that $\mathcal{C}_1$ is decaying exponentially as a function of $\mu$. However the decay of $\mathcal{C}_2$ is not as fast as exponential. }
\end{figure}

The indefinite integration inside the absolute value takes the form
\begin{align}\label{eq:c1indefinite}
&      \int dx_1 ~
    x_1^{-1/2}H_{i\mu}^{(1)}(x_1)e^{ix_1}
\nonumber\\
=~ & \frac{2^{1+i\mu}~x_1^{\frac{1}{2}-i\mu}}{\pi(i+2\mu)}~\Gamma(i\mu)~
      {}_2F_2\left(\frac{1}{2}-i\mu,\frac{1}{2}-i\mu;
      \frac{3}{2}-i\mu,1-2i\mu; 2ix_1\right)
\nonumber\\
+~& e^{\pi\mu}~\frac{2^{1-i\mu}~x_1^{\frac{1}{2}+i\mu}}{\pi(i-2\mu)}~\Gamma(-i\mu)~
      {}_2F_2\left(\frac{1}{2}+i\mu,\frac{1}{2}+i\mu;
      \frac{3}{2}+i\mu,1+2i\mu; 2ix_1\right)~,
\end{align}
where ${}_2F_2(a_1,a_2;b_1,b_2;z)$ is the hyper-geometric function defined as
\begin{align}
  {}_2F_2(a_1,a_2;b_1,b_2;z) \equiv \sum_{n=0}^{\infty}
\frac{a_1(a_1+1)\cdots(a_1+n-1)~ a_2(a_2+1)\cdots(a_2+n-1)}
{b_1(b_1+1)\cdots(b_1+n-1)~ b_2(b_2+1)\cdots(b_2+n-1)} ~\frac{z^n}{n!}~.
\end{align}
The $x_1=0$ limit of equation (\ref{eq:c1indefinite}) vanishes. In the $x_1\rightarrow +\infty$ limit, using the asymptotic behavior
\bea
&& {}_2F_2(a,a;b_1,b_2;z) \xrightarrow{|z|\to \infty} \frac{\Gamma(b_1)\Gamma(b_2)}{\Gamma(a)^2} e^z z^{2a-b_1-b_2}
\nonumber \\
&& + \frac{\Gamma(b_1) \Gamma(b_2)}{\Gamma(a) \Gamma(b_1-a) \Gamma(b_2-a)}
(-z)^{-a}
\left( \ln(-z) - \psi(b_1-a) - \psi(b_2-a) - \psi(a) - 2\gamma \right) ~,
\eea
where
\bea
\psi(z) \equiv \frac{d\ln\Gamma(z)}{dz} ~,
\eea
Eq.~(\ref{eq:c1indefinite}) takes the form
\begin{align}
  \frac{\sqrt{\pi}(1-i)e^{\frac{\pi\mu}{2}}}{\cosh(\pi\mu)}~,
\end{align}
where the oscillatory term $e^{2ix_1}$ at infinity $x_1\to \infty$ is take to be zero by an $i\epsilon$-description.
Taking the absolute value and plugging it in (\ref{eq:c1}), one finds
\begin{align}
  \mathcal{C}_1 = \frac{\pi^2}{4\cosh^2(\pi\mu)}~.
\label{eq:resultc1}
\end{align}
A plot of $\mathcal{C}_1$ as a function of real $\mu$ is shown in Fig.~\ref{fig:num}. We also numerically check the integration and plot it in the same figure. The method of numerical calculation is described in Appendix \ref{App:Numerical}.

In the large $M/H$ limit, the above result goes as
\begin{align}
  \lim_{\mu\rightarrow\infty} \mathcal{C}_1 = \pi^2 e^{-2\pi M/H}~.
\end{align}
Interestingly, the suppression in $\mathcal{C}_1$ is a Boltzmann factor. However, as we shall show in the next section, this Boltzmann factor is not the dominate contribution to the two-point function.

\section{The integral with time-ordering}
\label{sec:c2}
\setcounter{equation}{0}

The integral with the time-ordering and double integration is the more difficult one,
\begin{align}\label{eq:c2}
  \mathcal{C}_2 \equiv -\frac{\pi}{4} ~ e^{-\mu\pi}~\mathrm{Re}
  \left\{
    \int_0^\infty dx_1 ~
      x_1^{-1/2}H_{i\mu}^{(1)}(x_1)e^{-ix_1} \int_{x_1}^\infty dx_2 ~
      x_2^{-1/2}\left[H_{i\mu}^{(1)}(x_2)\right]^* e^{-ix_2}
  \right\}~.
\end{align}
The inner layer of (\ref{eq:c2}) can be integrated out as
\begin{align}
  \mathcal{I}(x_1) & \equiv \int_{x_1}^\infty dx_2 ~
      x_2^{-1/2}\left[H_{i\mu}^{(1)}(x_2)\right]^* e^{-ix_2}
\label{2nd_integral_layer1}
      \\ &
      = \sqrt{\pi}(1+i)e^{\frac{\mu\pi}{2}}\mathrm{sech}(\mu\pi)
      \nonumber\\ &
      + \frac{2^{1+i\mu}e^{\pi\mu}x_1^{\frac{1}{2}-i\mu}\Gamma(i\mu)}{\pi(i+2\mu)}
      ~{}_2F_2\left(\frac{1}{2}-i\mu,\frac{1}{2}-i\mu;
      \frac{3}{2}-i\mu,1-2i\mu; -2ix_1\right)
      \nonumber\\ &
      + \frac{2^{1-i\mu}x_1^{\frac{1}{2}+i\mu}\Gamma(-i\mu)}{\pi(i-2\mu)}
      ~{}_2F_2\left(\frac{1}{2}+i\mu,\frac{1}{2}+i\mu;
      \frac{3}{2}+i\mu,1+2i\mu; -2ix_1\right)~.
\label{2nd_integral_layer1_result}
\end{align}
The first term above is the $x\to \infty$ limit of the indefinite integral, and the other two terms come from the $x\to x_1$ side.

The difficult part is the integration in the second layer,
\begin{align}
  \int_0^{\infty} dx_1 ~
      x_1^{-1/2}H_{i\mu}^{(1)}(x_1)e^{-ix_1} \mathcal{I}(x_1) ~.
\label{eq:outter}
\end{align}
With both the special function ${}_2F_2$ and $H^{(1)}_{i\mu}$ in the integrand, the indefinite integral cannot be done by Mathematica 8. We use a trick of \textit{resummation}. We expand the part of the integrand in the second layer, $x_1^{-1/2}H_{i\mu}^{(1)}(x_1)e^{-ix_1}$, in a series expansion in bases of $x^{n\pm i\mu -1/2}$. With the Hankel function replaced by the simple power-law, the definite integration can be done for each term in this series expansion. And it turns out that we can finally re-sum this series and get a closed-form result.

To do this we rewrite
\begin{align}
  H^{(1)}_{i\mu}(z) = J_{i\mu}(z) + i Y_{i\mu}(z)
  = [1+\coth(\pi \mu)] J_{i\mu}(z) - \frac{J_{-i\mu}(z)}{\sinh(\pi \mu)}~.
\end{align}
The Bessel $J_{i\mu}$ function, together with some $x_1$-dependent coefficients, can be expanded as
\begin{align}
  x_1^{-1/2}J_{i\mu}^{(1)}(x_1)e^{-ix_1} = \sum_{n=0}^{\infty} c^+_n(x_1)~,\qquad
  c^+_n(x_1)\equiv \frac{2^{n+i\mu}(-i)^n x_1^{n+i\mu-\frac{1}{2}}\Gamma(n+\frac{1}{2}+i\mu)}
  {n!\sqrt{\pi}\Gamma(n+1+2i\mu)}~.
\end{align}
As noted, $c_n(x_1)$ is proportional to a simple power of $x_1$, thus can be integrate out in the $x_1$ integration. The result is \footnote{Direct calculation, after an asymptotic expansion of the generalized hyper-geometric functions, shows $$\mathcal{J}_n^+ = -\frac{4 e^{(\mu-in)\pi}}{\pi(2\mu-i-2in)^2} + (f_n^{(1/2+n,+)} x_1^{1/2 +n + i\mu} + f_n^{(-1/2+n, +)} x_1^{-1/2 + n + i\mu}+ \cdots + f_n^{(1/2, +)} x_1^{1/2 + i\mu} )|_{x_1\rightarrow\infty} ~,$$ where $f_n^{(\cdots)}$ are independent of $x_1$.
Note that, in (\ref{2nd_integral_layer1_result}), the hyper-geometric function contains an oscillation factor $e^{-2ix_1}$ at $x_1\to \infty$. This behavior makes $\CJ^+_n$ either convergent or contain the similar oscillatory behavior, after we evaluate the result of the indefinite integral at $x_1\to \infty$. The above expansion contains finite number of terms for each $n$ and does not have this kind of oscillatory behavior. So it must be finite. This shows that all $f_n^{\cdots}$ must vanish, leaving only the first finite term. Explicitly we have only checked two of them, $f_n^{(1/2+n, +)}$ and $f_n^{(-1/2+n, +)}$, indeed vanish. Similar conclusion holds for $\CJ^-_n$ later.}
\footnote{It is useful to deform the factor $x_1^{-1/2}$ in (\ref{eq:outter}) to $x_1^{-1/2+\alpha}$ by a small number $\alpha$. This avoids some spurious divergences coming from some hypergeometric and Gamma functions in the intermediate steps. After the integration is done, we can take $\alpha\to 0$ limit and get a regular answer.}
\begin{align}
\mathcal{J}^+_n \equiv  \int_0^\infty dx_1 ~ \mathcal{I}(x_1) c^+_n(x_1)
=-\frac{4 (-1)^n e^{\mu\pi}}{\pi(2\mu-i-2in)^2}~.
\end{align}
The function $\mathcal{J}^+_n$ can be re-summed with respect to $n$,
\begin{align}
  \sum_{n=0}^{\infty}\mathcal{J}^+_n = \frac{e^{\mu\pi}}{4\pi}
    \left[
      \psi^{(1)}\left(\frac{1}{4}+\frac{i\mu}{2}\right)
      -\psi^{(1)}\left(\frac{3}{4}+\frac{i\mu}{2}\right)
    \right]~,
\end{align}
where
\begin{align}
  \psi^{(1)}(z)\equiv \frac{d}{dz}\left(\frac{d\Gamma(z)/dz}{\Gamma(z)}\right)~.
\end{align}
Similarly, the Bessel $J_{-i\mu}$ function, together with its $x_1$-dependent coefficients, can be expanded as
\begin{align}
  x_1^{-1/2}J_{-i\mu}^{(1)}(x_1)e^{-ix_1} = \sum_{n=0}^{\infty} c^-_n(x_1)~,\qquad
  c^-_n(x_1)\equiv \frac{2^{n-i\mu}(-i)^n x_1^{n-i\mu-\frac{1}{2}}\Gamma(n+\frac{1}{2}-i\mu)}
  {n!\sqrt{\pi}\Gamma(n+1-2i\mu)}~.
\end{align}
The $x_1$ integration can be performed term by term in $n$ as
\begin{align}
\mathcal{J}^-_n \equiv  \int_0^\infty dx_1 ~ \mathcal{I}(x_1) c^-_n(x_1)
=-\frac{4 (-1)^n}{\pi(2\mu+i+2in)^2}~.
\end{align}
We re-sum over $n$ and get
\begin{align}
  \sum_{n=0}^{\infty}\mathcal{J}^-_n = \frac{1}{4\pi}
    \left[
      \psi^{(1)}\left(\frac{1}{4}-\frac{i\mu}{2}\right)
      -\psi^{(1)}\left(\frac{3}{4}-\frac{i\mu}{2}\right)
    \right]~.
\end{align}

Combining the results we get
\begin{align}\label{eq:resultc2}
  \mathcal{C}_2 = & -\frac{\pi}{4} ~ e^{-\mu\pi}~\mathrm{Re} \left\{
\left[1+\mathrm{coth}(\pi\mu)\right]\mathcal{J}^+_n - \frac{\mathcal{J}^-_n}{\sinh\pi\mu}\right\}
\nonumber\\
=&
\frac{e^{\pi\mu}}{16 \sinh\pi\mu}
    \mathrm{Re}\left[
      \psi^{(1)}\left(\frac{3}{4}+\frac{i\mu}{2}\right)
      -\psi^{(1)}\left(\frac{1}{4}+\frac{i\mu}{2}\right)
    \right]
\nonumber\\
-&\frac{e^{-\pi\mu}}{16 \sinh\pi\mu}
    \mathrm{Re}\left[
      \psi^{(1)}\left(\frac{3}{4}-\frac{i\mu}{2}\right)
      -\psi^{(1)}\left(\frac{1}{4}-\frac{i\mu}{2}\right)
    \right]~.
\end{align}
A plot of $\mathcal{C}_2$ is shown in Fig.~\ref{fig:num}, again together with the numerical check.

In the large $M/H$ limit, the second line of (\ref{eq:resultc2}) is exponentially suppressed, and the first line is power-law suppressed thus dominates. As a result,
\begin{align}
  \lim_{\mu\rightarrow\infty}\mathcal{C}_2 = \frac{H^2}{4M^2}~.
\label{Eq:C2_limit}
\end{align}
Thus there is no Boltzmann factor in $\mathcal{C}_2$. In the large $M$ limit, $\mathcal{C}_2$ is much greater than $\mathcal{C}_1$, and is the dominate contribution in this limit.

\section{Analytical continuation to imaginary $\mu$}
\label{Sec:Analytical_continuation}
\setcounter{equation}{0}

\begin{figure}[htbp]
\centering
\includegraphics[width=0.6\textwidth]{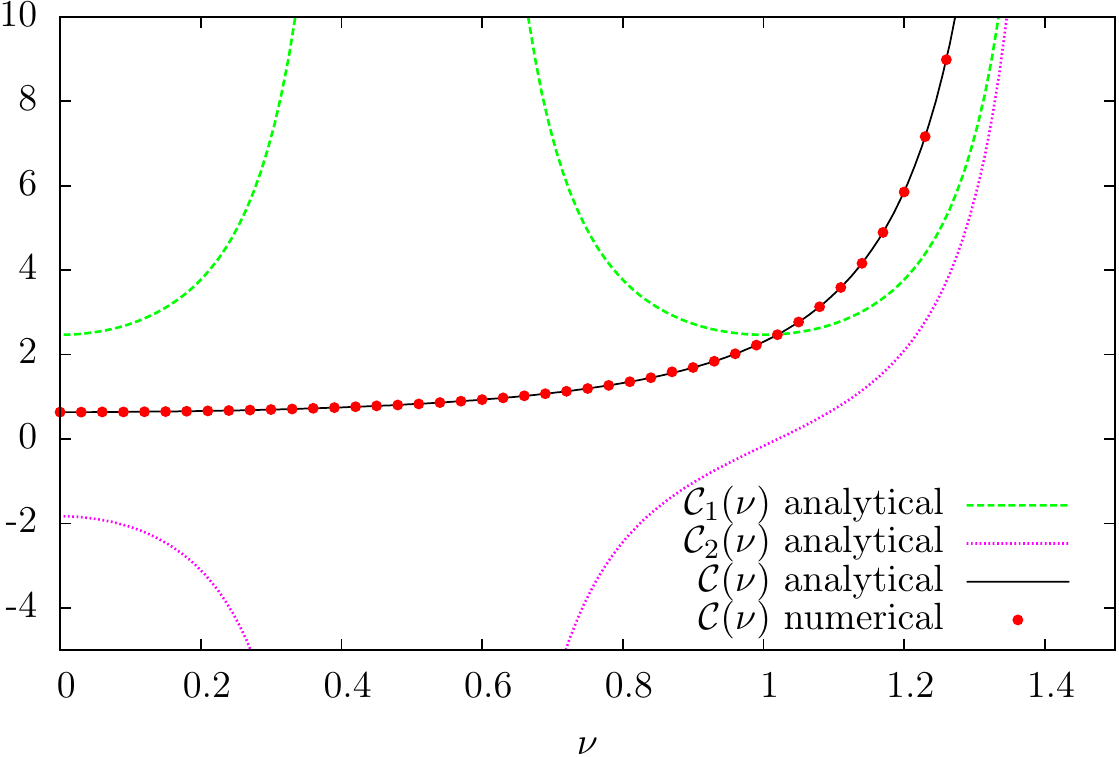}
\caption{\label{fig:real-crop} The analytical continuation of $\CC_1(\mu)$ and $\CC_2(\mu)$ to imaginary $\mu$ (corresponds to real $0<\nu<3/2)$. The solid line is $\CC=\CC_1+\CC_2$. The dots are numerical results from \cite{Chen:2009zp}.}
\end{figure}

In this section we extend our results to the case of lighter isocurvaton, $0\le M <3H/2$, by analytically continuing $\mu \to - i \nu$. As we have seen in \cite{Chen:2009we,Chen:2009zp}, both $\CC_1$ and $\CC_2$ should have spurious divergences at $\nu=1/2$, but those divergences cancel each other. These properties precisely show up after the analytical continuation of (\ref{eq:resultc1}) and (\ref{eq:resultc2}), because $\cosh^{-2}(x)$ in (\ref{eq:resultc1}) has a pole at $x=-i\pi/2$ and $\psi^{(1)}(x)$ in (\ref{eq:resultc2}) has a pole at $x=0$.
The final result also matches well with the numerical results in Fig.~6 of Ref.~\cite{Chen:2009zp}. These are shown in Fig.~\ref{fig:real-crop} here. This confirmation provides a non-trivial check that we indeed have the exact results.

\section{An IR expansion for large $\mu$}
\label{Sec:IR_Approx}
\setcounter{equation}{0}

In the previous sections, we have obtained the power spectrum correction from isocurvaton with arbitrary mass in quasi-single field inflation. However, if one is interested in more general scenarios than the present model, the exact method in the above sections could become very difficult or impossible to apply. Thus in this section, we present an approximation method -- the \textit{IR expansion method}. This method is applicable \textit{only} in the large $\mu$ limit, \textit{and} when the results are power-law (instead of exponential) suppressed, as we will explain. But on the other hand it involves simpler special functions and may be more friendly to be applied in more complicated cases.

In the small argument limit, the Hankel function can be expanded as
\bea
H^{(1)}_{i\mu}(x) \to \frac{1+\coth(\pi\mu)}{\Gamma(1+i\mu)} \left( \frac{x}{2} \right)^{i\mu} - i \frac{\Gamma(i\mu)}{\pi} \left( \frac{x}{2} \right)^{-i\mu} ~.
\label{Hankelsmallx}
\eea
For small $x$, the expansion parameter for the Hankel function is $\sim \frac{x^2}{4(n+1+\mu)}$ for terms of order $\CO(x^{2n\pm i\mu})$. So this leading order approximation is only good for $0<|x|<\sqrt{\mu}$. We emphasize that physically this is more than a good approximation in the IR limit. It includes the behaviour of the massive mode function at the super-Hubble limit $k\tau \to 0$, as well as an important portion of sub-Hubble scales $-\sqrt{M/H}< k\tau <0$. The behavior for larger $x$,
\bea
H^{(1)}_{i\mu}(x) \to -(-1)^{3/4}\sqrt{\frac{2}{\pi x}} e^{\pi\mu/2} e^{ix} ~,
\quad x>\mu^2 ~,
\label{Hankellargex}
\eea
is not well represented at all by (\ref{Hankelsmallx}). The approximation method here is to replace the correct deep UV behavior in (\ref{Hankellargex}) with the UV behavior in (\ref{Hankelsmallx}), and
run the integrals from $0$ to $\infty$ with the approximation (\ref{Hankelsmallx}).

To use this method, we need to satisfy the following two requirements for the integrals of interest. After a Wick rotation $x=\pm iy$ (which is explained below),
we refer to $y<\sqrt{\mu}$ as the IR region, and $y>\sqrt{\mu}$ as the UV region.
First,  the UV behavior of (\ref{Hankelsmallx}) only contributes to the final result a term that is exponentially suppressed by large $\mu$; second, the correct UV behavior of the Hankel function also gives an exponentially small contribution. Then, if this approximation method gives a result that is power-law suppressed (instead of exponentially suppressed) by large $\mu$, this result is the correct leading order approximation.

We demonstrate how this method works for the two integrals that we computed previously.
We first estimate the integral (\ref{eq:c1}), which is schematically
\bea
\sim e^{-\pi\mu}
\bigg|\int_0^\infty dy ~ y^{-1/2} H^{(1)}_{i\mu}(iy) e^{-y} \bigg|^2 ~.
\label{int_est_1}
\eea
Here it is important to note that we have made a Wick rotation $x=iy$. This rotation turns the oscillatory factor $e^{ix}$ into the exponential decay factor. Otherwise the UV contributions are not under as good control in the below estimates.

First, we look at the UV contribution from the IR expansion method.
In the large $\mu$ limit, (\ref{Hankelsmallx}) is schematically
\bea
H^{(1)}_{i\mu}(iy) \sim y^{i\mu} + y^{-i\mu} ~,\qquad
H^{(1)}_{i\mu}(-iy) \sim e^{\pi \mu} y^{i \mu} + e^{-\pi \mu} y^{-i \mu}.
\label{H1_schematic}
\eea
Note that
\bea
\int_0^\infty dy~ y^{-1/2\pm i\mu} e^{-y} &\sim& e^{-\pi\mu/2} ~.
\eea
So for $\sqrt{\mu}<y<\infty$, the integral (\ref{int_est_1}) with the replacement (\ref{H1_schematic}) contributes no more than
\bea
e^{-\pi\mu} \bigg| e^{-\pi\mu/2} + e^{-\pi\mu/2} \bigg|^2
\sim e^{-2\pi\mu} ~.
\eea
Second, we estimate the actual UV contribution using (\ref{Hankellargex}). The approximation (\ref{Hankellargex}) is good for $y>\mu^2$, but can be used for $y>\sqrt{\mu}$ for our purpose of estimating the order of magnitude. The exponential factor $e^{-2y}$ in the integrand gives the suppression factor $e^{-2\sqrt{\mu}}$. So the actual UV contribution is of order
\bea
\sim e^{-4\sqrt{\mu}} ~.
\eea
Finally, using the IR expansion method, plugging (\ref{Hankelsmallx}) in (\ref{int_est_1}) and performing the resulting integration precisely, we get
\bea
8(1+\sin(2\mu\ln2))\frac{1}{\mu} e^{-2 \pi\mu}
\eea
in the large $\mu$ limit.

To summarize the situation so far, we see that, although the UV contribution from both the IR expansion method and the actual integration are exponentially suppressed by large $\mu$, the IR expansion method itself also gives an exponentially suppressed result. So we cannot get the exact leading order result using this method.
However, we do know from the above estimates that $\mathcal{C}_1$ is exponentially suppressed. Thus it is negligible comparing with $\mathcal{C}_2$, as we have shown in previous sections, and shall show again below.

We now look at the more important integral (\ref{eq:c2}), which is schematically
\bea
\sim
e^{-\pi\mu} \int_0^\infty dy_1~ y_1^{-1/2} H^{(1)}_{i\mu}(-iy_1)~ e^{-y_1}
\int_{y_1}^\infty dy_2~ y_2^{-1/2} H^{(2)}_{-i\mu}(-iy_2)~ e^{-y_2} ~.
\label{int_est_2}
\eea
Again here we have Wick-rotated such that $x_1=-iy_1$, $x_2=-iy_2$. The direction of Wick rotation here is different from that we have used for $\mathcal{C}_1$. This is because we need to keep the exponential factor to be $e^{-y_1}$ and $e^{-y_2}$, specified by the $i\epsilon$-prescription of the vacuum choice in the interacting field theory.

First, we estimate the UV contribution from the IR expansion method. After expanding the Hankel function using (\ref{Hankelsmallx}), for $y_1 > \mu$ the inner integration gives, schematically,
\bea
\sim \left( y_1^{i\mu}+y_1^{-i\mu} \right)
y_1^{-1/2}e^{-y_1} ~.
\eea
The above estimation can be extended to the region $y \gtrsim \sqrt{\mu}$ for our purpose. Combined with the integrand in the outer layer, the leading UV contribution is
\bea
\sim e^{-\pi\mu} \int_{\sqrt{\mu}}^\infty  dy_1~ y_1^{-1} e^{\pi\mu}  e^{-2y_1}
\sim e^{-2\sqrt{\mu}}
~.
\eea
Second, we estimate the actual UV contribution. Similarly as we did previously, we use (\ref{Hankellargex}) and extend it to $y>\sqrt{\mu}$ for our order of magnitude purpose. Due to the exponentially suppressed factor, the UV contribution is of order
\bea
\sim e^{-2\sqrt{\mu}} ~.
\eea

To summarize, we see again that the UV contributions from both the IR expansion method and the actual integration are exponentially suppressed. In the following we will precisely compute the entire integration (\ref{eq:c2}) using the IR expansion method (\ref{Hankelsmallx}), and show that it gives a power-law suppressed, hence the correct leading order result.

We replace both Hankel functions in (\ref{eq:c2}) with (\ref{Hankelsmallx}).
The inner layer can be directly computed and gives
\bea
(-1)^{1/4}(2i)^{-i\mu}
\left(
\frac{2^{2i\mu}{\rm csch}(\pi\mu) \Gamma(\half-i\mu,y_1)}{\mu \Gamma(-i\mu)}
+
\frac{\Gamma(-i\mu)\Gamma(\half+i\mu,y_1)}{\pi}
\right) ~.
\label{int2_inner}
\eea
However, combined with the integrand in the outer layer, the indefinite integral cannot be performed by Mathematica 8. We again use the trick of resummation -- expanding the outer integrand in terms of power series, performing the simpler integration with the power-law terms, and re-summing the series. The power-law expansion of the integrand in the outer layer is
\bea
\sum_{n=0}^{\infty} \left[
(-1)^{1/4}(2i)^{-i\mu} \frac{1+\coth(\pi\mu)}{\Gamma(1+i\mu)}  \frac{(-1)^n}{n!} y_1^{-1/2+i\mu+n}
+
(-1)^{-1/4}(2i)^{i\mu} \frac{\Gamma(i\mu)}{\pi}  \frac{(-1)^n}{n!}  y_1^{-1/2-i\mu+n}
\right] ~.
\label{int2_outer_expand}
\eea
With the power-law, the final indefinite integral is doable using the following formulae,
\bea
\int dx~ x^{-1+\alpha} {\rm E}_\nu(a x)
= \frac{x^\alpha}{\alpha+\nu-1} \left[
{\rm E}_\nu(ax) - {\rm E}_{1-\alpha}(ax) \right] ~,
\eea
where
\bea
{\rm E}_\nu(z) \equiv \int_1^\infty \frac{e^{-zt}}{t^\nu} dt = z^{\nu-1}\Gamma(1-\nu,z)
\eea
is the exponential integral function.
The resummation using the coefficients in (\ref{int2_outer_expand}) can be done by Mathematica. The final result is long and we do not list it here. We are only interested in the large $\mu$ limit, which is
\bea
\frac{1}{4\mu^2} ~,
\eea
the same as (\ref{Eq:C2_limit}).

\section{Truncating the interaction history}
\label{Sec:Truncating}
\setcounter{equation}{0}

As we have seen in the case of $0\le M/H \le 3/2$ \cite{Chen:2009we,Chen:2009zp}, the massive mode contributes to the curvature perturbation at different stages in its evolutionary history, sensitively depending on its mass. The lighter the isocurvaton is, the longer the contribution lasts after Hubble crossing. This is crucial in determining the shape of non-Gaussianities.

To investigate where the dominant contribution comes from for the heavier mode $M/H > 3/2$, we fix the upper limit of the integrals (\ref{eq:c1}) and (\ref{eq:c2}) to be $\infty$ and vary the lower limit $x_c$.\footnote{Note that the IR cutoff $x_c$ is different from the cutoff we considered in Sec. \ref{Sec:IR_Approx}. In Sec. \ref{Sec:IR_Approx}, the cutoff (as a separation between UV and IR regimes) is taken after performing a Wick rotation. Thus it is completely a mathematical trick, which does not change the final result of the integration but does change the intermediate behavior of the integration.
While here, we impose a cutoff in real time instead of imaginary time. Thus the cutoff $x_c$ here has the physical meaning as the termination time of interactions.
To keep the $x_c$ as the real-time cutoff, we can only Wick-rotate {\em after} shifting $x\to x + x_c$, as we describe in Appendix \ref{App:Numerical}.}

There are at least two purposes for doing this: First, it is interesting to know the conversion from isocurvature to curvature perturbations is dominated by which period of time. Second, one could imagine models in which the turning of trajectory ends at some time before reheating. We hope this truncation of interaction can illustrate part of this effect.\footnote{When the turning of trajectory ends, the physical process could be complicated: (1) the transfer from isocurvature to curvature terminates; (2) there may be classical oscillation because of the inertia of the inflaton; (3) the potential which stops the turning of trajectory may have extra features; (4) part of the vacuum state of the isocurvature direction is projected into a non-vacuum state of the inflaton direction; $\dots$. Here by putting a cutoff $x_c$, we only included (1). Nevertheless this contributes part of the observable effects. Moreover, (2) and (3) could be eliminated by fine-tuning parameters in the potential, in which cases those effects may be separated.}

\begin{figure}[htbp]
\centering
\includegraphics[height=0.34\textheight]{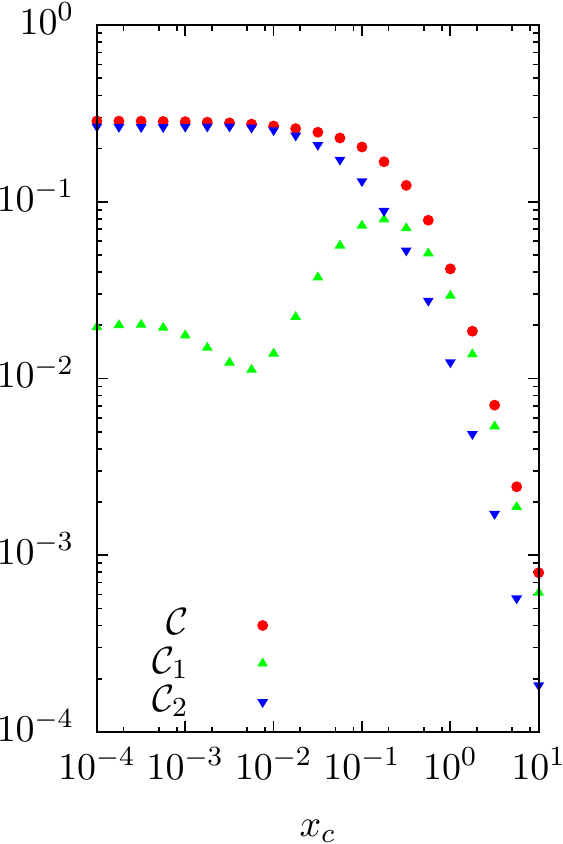}
\includegraphics[height=0.34\textheight]{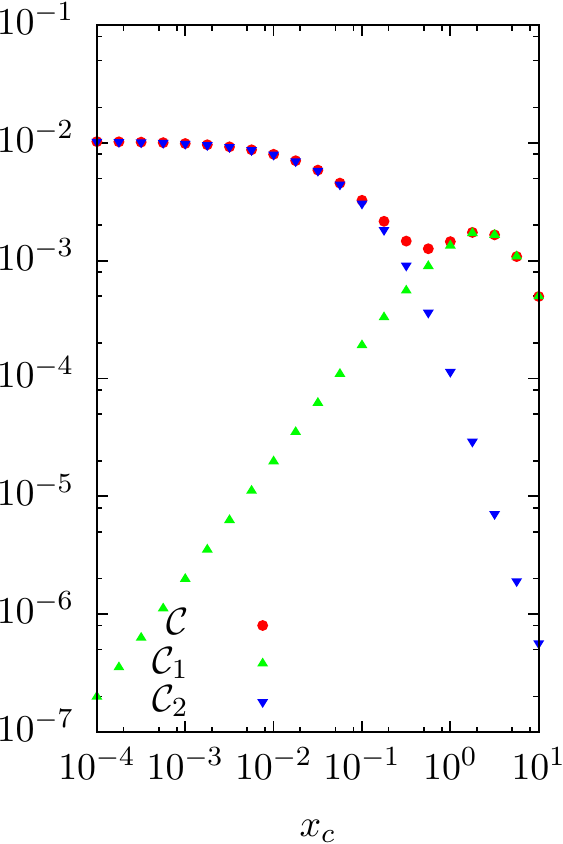}
\includegraphics[height=0.34\textheight]{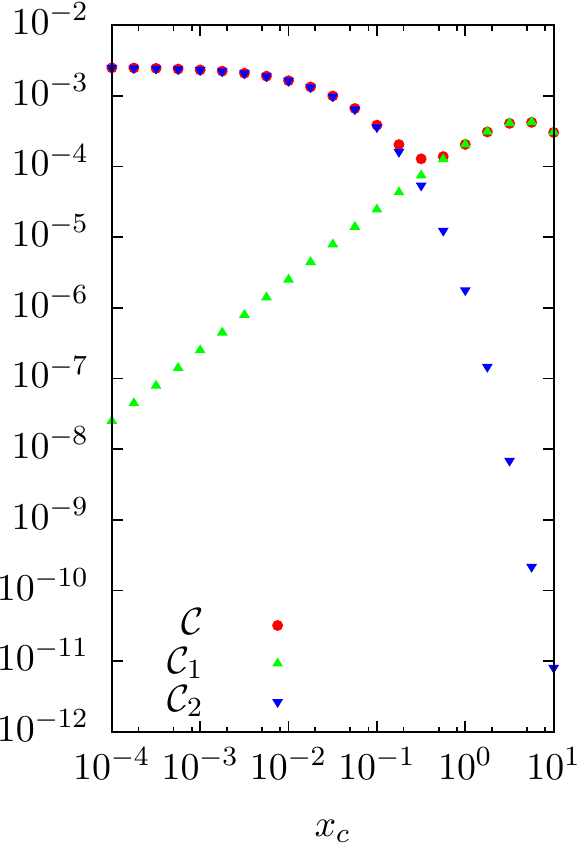}
\caption{\label{fig:contribuation}  The coefficients $\mathcal{C}$, $\mathcal{C}_1$ and $\mathcal{C}_2$ as a function of IR cutoff $x_c$. The left, middle, right panels are with $\mu=1$, $\mu=5$ and $\mu=10$ respectively.}
\end{figure}

We numerically compute the integrals as a function of $x_c$ for different $\mu$ and plot them in Fig.~\ref{fig:contribuation}. The numerical result shows that $\mathcal{C}_2$ decays quickly when we push $x_c$ into the far UV regime. On the other hand, it is interesting to see that the maximum UV contribution of $\mathcal{C}_1$ is actually also $\sim 1/\mu^2$ for large $\mu$, but the effect of $\CC_1$ decays a few e-folds after the Hubble crossing.
As a result, for large $\mu$, the net contribution of $\CC_1$ and $\CC_2$ does not grow monotonically in time (although the effects are small).
This (if other effects do not wash it away) is a counter example of a no-go theorem stating that the curvature perturbation cannot be suppressed by the isocurvature perturbation (see \cite{Sloth:2005yx} for a discussion of super-Hubble perturbations and \cite{Li:2008jn} for both sub- and super-Hubble). The no-go theorem is violated here because the isocurvature direction does not satisfy the slow-roll conditions, which is an assumption in \cite{Li:2008jn}.

In the case of a finite turn trajectory, Fig.~\ref{fig:contribuation} can also be reinterpreted as the contribution of the massive isocurvaton to the power spectrum as a function of comoving wave number $k$, with the replacement $x_c\rightarrow k/(a_c H)$. In this way the pattern in Fig.~\ref{fig:contribuation} is a contribution to the power spectrum as a function of $k$ due to the finite turn.

\begin{figure}[htbp]
\centering
\includegraphics[height=0.25\textheight]{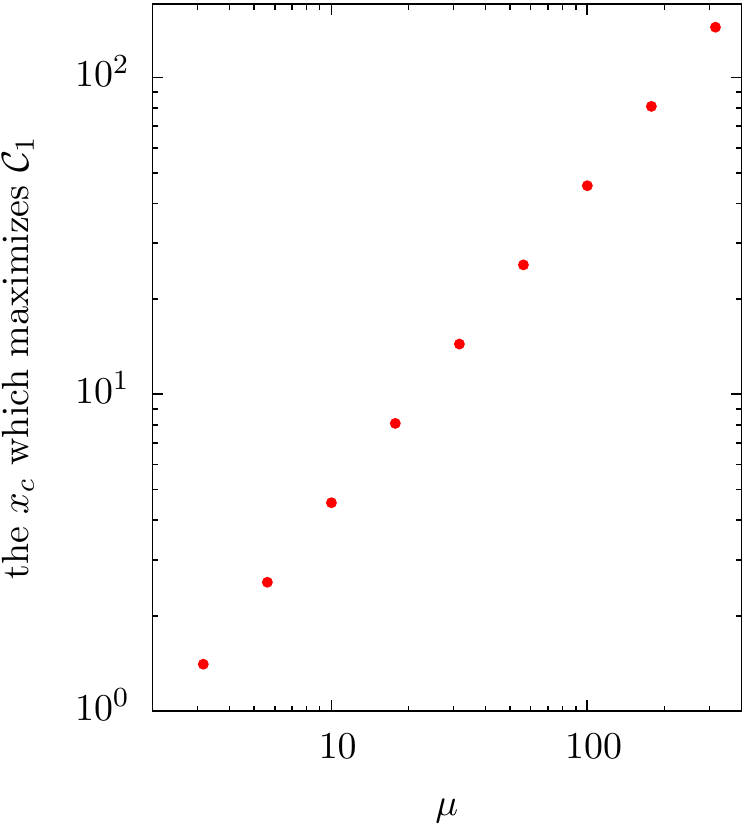}
\hspace{0.1\textwidth}
\includegraphics[height=0.25\textheight]{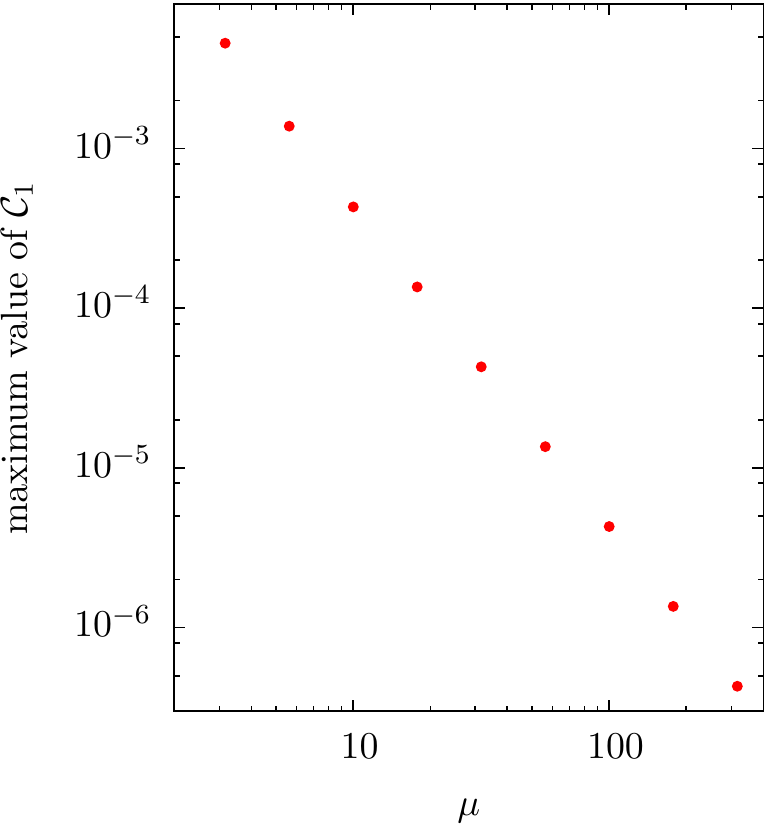}
\caption{\label{fig:max-val-and-pos}  When an IR cutoff $x_c$ is considered, $\mathcal{C}_1$ has a maximum when considered as a function of $x_c$ (see the plot of $\mathcal{C}_1$ in Fig.~\ref{fig:contribuation}). The left panel plots the peak position $x_c$ which maximizes $\mathcal{C}_1$, for different $\mu$. The right panel plots the corresponding maximum values of $\mathcal{C}_1$ for those $\mu$ values. For large $\mu$, one can read from the plot that the peak position is at $x_c \approx 0.5\mu$, where $\mathcal{C}_1 \approx 0.05/\mu^2$.}
\end{figure}

Overall, for the constant turn case we are interested in in the previous sections, during the whole evolution, the massive isocurvaton contributes most significantly during the time interval from of order $\ln\mu$ e-folds before the Hubble crossing, to a few e-folds after it. Several e-folds after the Hubble crossing, the amplitude of the massive mode decays away. However it is a special feature of the massive mode with $\mu\gg 1$ that starting from $\ln\mu$ e-folds before the Hubble crossing, the isocurvaton contribution already takes the similar order of magnitude as its eventual value. This feature can be observed in Fig.~\ref{fig:max-val-and-pos}.
The $\CC_1$, which is negligible after the Hubble-crossing, now becomes the dominate contribution before the Hubble-crossing, for $\mu \gg 1$. Using results in Sec.~\ref{sec:c1}, this term can be studied both numerically and analytically even with the cutoff $x_c$. It has a maximal value $\approx 0.05/\mu^2$, at $x_c \approx 0.5\mu$.

Because the dominant isocurvature contribution comes from within a few e-folds, the constant-turn assumption made so far is not essential for the main results of this paper. As long as the variation of various parameters satisfy the conditions of the slow-variation approximation, for example, $\ddot\theta/(H\dot\theta) \ll 1$ and $\dot M/(H M) \ll 1$, the same results apply with all the parameters take their instantaneous values.\\

Notes added: When this paper was in completion, we became aware of a related work \cite{Pi:2012gf}. We thank the authors for coordinating on the submission of both works.

\medskip
\section*{Acknowledgements}

XC is supported by the Stephen Hawking Advanced Fellowship. YW is supported by a fellowship from McGill University.

\appendix

\section{Numerical computation of the integrals}
\label{App:Numerical}
\setcounter{equation}{0}

Although we have got full analytic results in Sec.~\ref{sec:c1} \& \ref{sec:c2}, it is useful to check the calculation numerically. Note that the integrands in both integrals are oscillating at $x_1\rightarrow\infty$. The oscillation makes the integration difficult (though still possible) to perform. There are a few techniques available to deal with such oscillations. For example, one can insert an $e^{-\epsilon x_1}$ term, keeping $\epsilon$ to be a small number. Or one can integrate the expression by parts to obtain better convergence. Here we use a third method, as used in \cite{Chen:2009zp}: We Wick rotate $x_1$ and $x_2$ into imaginary values, such that $e^{\pm ix_1}$, $e^{\pm ix_2}$ $\rightarrow$ $e^{-x_1}$, $e^{-x_2}$. Here the directions of rotation are chosen to ensure the convergence of the integrals, corresponding to the $i\epsilon$ prescription in interacting field theories.

For example, in $\mathcal{C}_2$, we first shift $x_1 \to x_1 + x_c$, $x_2 \to x_2 + x_1$ and then rotate $x_1\rightarrow -iy_1$, $x_2\rightarrow -iy_2$. We have
\begin{align}
  & \int_{x_c}^{\infty}dx_1\int_{x_1}^{\infty}dx_2 ~ f(x_1,x_2)
  = \int_{0}^{\infty}dx_1\int_{0}^{\infty}dx_2 ~ f(x_1+x_c,x_2+x_1)
  \nonumber\\
  = & - \int_{0}^{\infty}dy_1\int_{0}^{\infty}dy_2 ~ f(-i y_1+x_c, -iy_2-iy_1)~.
\end{align}Here $f$ denotes the integrand in $\mathcal{C}_2$, and $x_c$ is an IR cutoff for the integral, added for two purposes: (1) In Fig.~\ref{fig:contribuation}, we have plotted $\mathcal{C}_2$ as a function of $x_c$. Note that in this case, the Wick-rotation after the shifts does not change the fact that $x_c$ is the real time.
(2) Even in the case without an IR cutoff, we actually also put a small $x_c$ in the above integral to prevent evaluating into 0/0 in the integrand. In the latter case we have checked that the $x_c$ we put ($x_c = 10^{-10}$) is small enough such that the result is stable against order-of-magnitude variation of $x_c$.

\end{spacing}

\end{document}